# Tunable bound states in the continuum through hybridization of 1D and 2D metasurfaces

Fedor Kovalev[1, 4, *], Mariusz Martyniuk[2, 4], Andrey Miroshnichenko[3], and Ilya Shadrivov[1, 4]

[1] Research School of Physics, The Australian National University, Canberra ACT 2601, Australia
[2] School of Engineering, The University of Western Australia, Perth, WA 6009, Australia
[3] School of Engineering and Technology, University of New South Wales Canberra, Canberra ACT 2612, Australia
[4] Australian Research Council Centre of Excellence for Transformative Meta-Optical Systems, Australia
* *fedor.kovalev@anu.edu.au*

**Abstract**

This work presents a novel approach to create and dynamically control quasi-Bound States in the Continuum (BIC) resonances through the hybridization of 1D and 2D metasurfaces using micro-electromechanical systems (MEMS). By introducing out-of-plane symmetry breaking through a silicon MEMS membrane positioned above a 1D silicon metasurface, the quasi-BIC resonance's central wavelength and quality factor are precisely tuned. The proposed design achieves ultranarrow resonance linewidths with the spectral tuning range exceeding 60 nm while maintaining a constant quality factor. This tuning capability, realized through both horizontal displacement within a 1D metasurface and vertical MEMS membrane movement, offers a new degree of freedom for manipulating quasi-BIC resonances. The proposed hybridization of 2D and 1D metasurfaces using MEMS mechanism provides a practical route to dynamic modulation of transmission resonance characteristics, making it a promising candidate for tunable filters, spectroscopy, imaging, and sensing applications.

**Introduction**

The concept of Bound States in the Continuum (BICs) represents a remarkable phenomenon in modern physics, where certain states remain localized within a system despite existing at energies or frequencies that typically allow for radiation [1]. Originating from quantum mechanics, BICs have found significant applications in photonics and metamaterials, enabling extreme confinement of electromagnetic waves and leading to the formation of extremely high-Q resonances [2,3]. These unique properties of BICs make them invaluable for enhancing light-matter interactions, paving the way for developing advanced photonic devices [4].

Recent advancements have focused on the active control of BICs [5], where external stimuli, such as electric fields, temperature variations, or material phase transitions, are used to dynamically tune their properties [6]. This capability enables real-time reconfiguration [7], opening up opportunities for a wide range of applications, including tunable filters [8,9], sensors [10–12], and modulators [13–15], which are critical components in modern communication, imaging, and sensing technologies. The precise manipulation of quasi-BIC (q-BIC) resonances in real-time could revolutionize photonic systems, enhancing device performance and functionality, and making this a rapidly growing field in applied physics and engineering [16].

Micro- and nano-electromechanical systems (MEMS/NEMS) have emerged as a leading technology in fields such as active nanophotonics, sensing, and telecommunications, due to their compactness, precise control, and low power consumption [17–20]. These systems offer dynamic, strong on-demand tuning of optical properties, which has revolutionized many applications by enabling integration into smaller, more efficient devices. One of the most significant breakthroughs in this area is the control of high-Q resonances in metasurfaces through mechanical actuation [21–23]. This has led to significant advancements, particularly in imaging and active photonics, allowing for the real-time manipulation of optical properties with minimal energy input [24,25].

Recently, the control of guided-mode resonances (GMR) using MEMS/NEMS has opened new avenues for dynamic manipulation of light in nanophotonic devices [26–29]. By integrating MEMS/NEMS, the mechanical actuation of these structures allows for precise, real-time tuning of resonance characteristics, enabling modulation of amplitude, phase, and wavelength. Notably, NEMS technology has been employed to tune both GMR and q-BIC resonances in all-dielectric metasurfaces, enabling dynamic modulation of both amplitude and phase [30]. The interference between these two resonant modes significantly enhanced the phase response of the metasurfaces, allowing for precise control with minimal input. In particular, a phase shift of 144° was achieved experimentally with just a 4 V bias change. However, this approach exhibited non-zero reflection at the q-BIC resonance, which is less than ideal for imaging applications that require minimal reflection for optimal performance. Exploring the q-BIC resonances in transmission is critical for applications such as spectroscopy, sensing and imaging, where high transmission and precise control of resonances are essential for accurate data acquisition [19].

More recently, static and thermally tunable BIC-based metasurfaces have been proposed for use in spectrometers, achieving sub-nm resolution [31,32]. While the proposed static designs in the optical range rely on bandstop filtering and require multiple metasurfaces to cover the desired spectral range, leading to extra computations and increased spectrometer size, respectively; the designed ultracompact tunable solutions in the infrared range face limitations, including the inability to maintain a constant bandwidth and a relatively narrow spectral tuning range of only 6 nm.

In this work, we introduce MEMS-enabled comprehensive control over q-BIC resonances in a transmission configuration within all-silicon metasurfaces. The q-BIC resonance is achieved via breaking the out-of-plane symmetry by combining a 2D MEMS structure that moves close to a 1D metasurface without direct contact. We demonstrate that the resonance quality factor approaches infinity as the MEMS structure moves farther from a 1D metasurface, thereby reducing the asymmetry - a defining characteristic of q-BIC resonances. The vertical movement of a MEMS structure combined with horizontal displacement within a 1D metasurface enables the central wavelength to be shifted by more than 60 nm while maintaining a consistent resonance bandwidth and Q-factor. The proposed all-silicon metasurfaces achieve an ultranarrow resonance linewidth with dynamic tunability in the infrared range, making them promising candidates for advanced tunable filtering applications and highly relevant for spectroscopy, remote sensing, and imaging.

**Results**

We begin by investigating high-Q q-BIC resonances excited in all-silicon metasurfaces by a plane wave under normal incidence. Figure 1a presents a conceptual design of a resonant structure composed of a 1D metasurface made of silicon bars and an overlying 2D array of cuboids. The structural parameters shown in Figure 1a are as follows: $w_c$ = 0.8 µm, $h_c$ = 1 µm, $w_r$ = 0.8 µm, $h_r$ = 0.5 µm, $d$ = 0.7 µm. We model the transmission in CST Studio, where we employ Floquet ports and periodic boundary conditions. The edge length of the square unit cell ($p$) is 1.5 µm. The polarization of the incident plane wave is shown in Figure 1a.

Figure 1b illustrates the electric field distribution at the q-BIC resonance when the distance between the cuboids and bars ($a$) is 1 µm. Transmission spectra for various distances between the bars and cuboids are depicted in Figure 1c. This structure enables out-of-band transmission of near than or below 0.1 within the 3100–3300 nm range when tuning $a$ from 0.1 to 1 µm and less than 0.04 across 2700–3400 nm when $a$ is set to 1 µm.

In the absence of cuboids, no q-BIC resonance is observed, as the non-radiating BIC mode remains confined within the bars. Nevertheless, the BIC mode was identified using the eigenmode solver. Introducing asymmetry through the positioning of the cuboids above the bars enables radiation leakage, leading to the observation of high-Q resonances in transmission.

We use the Fano formula [33,34] to fit the transmission curves shown in Figure 1c, enabling accurate analysis of the characteristics of the excited q-BIC resonances while accounting for their asymmetry:

$$T_{Fano}(\omega) = t_{d_0} + t_d^2 \frac{\left[q + \frac{2(\omega - \omega_0)}{d\omega}\right]^2}{1 + \frac{4(\omega - \omega_0)^2}{d\omega^2}}, \tag{1}$$

where $t_{d_0}$ is the minimum of transmission near resonance, $t_d$ is the non-resonant transmission amplitude, $q$ is the shape factor, $\omega_0$ is the resonance frequency, $d\omega$ is the resonance width.

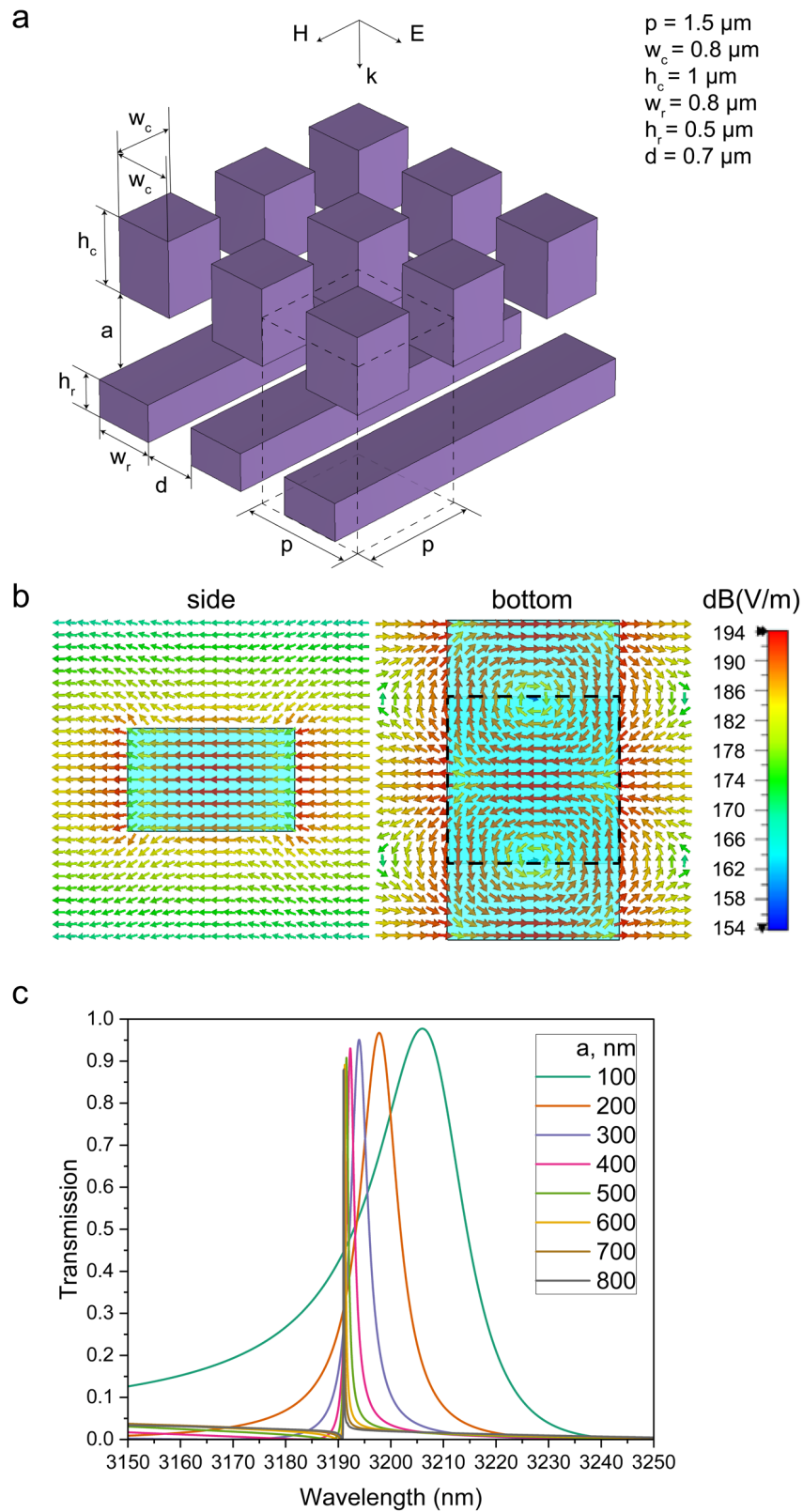

Figure 1. a) Schematic of the tunable metasurface conceptual design: silicon bars and an array of cuboids, including the unit cell indicated by the dashed lines; b) Maximum electric field amplitude of the BIC mode within the bar for $a$ = 1 μm (side and bottom views, the dashed lines show the position of the cube); c) Transmission coefficient for various distances between silicon bars and cuboids ($a$).

Figure 2 a and b shows the dependence of the q-BIC resonance wavelength and the Q-factor on the distance between the bars and the cuboids. The q-BIC resonance wavelength approaches the BIC value of approximately 3193.2 nm as the cuboids move away from the 1D metasurface. To accurately describe the relationship between the Q-factor of the excited q-BIC resonances and the distance between the cuboids and the bars depicted in Figure 2b, we can use the exponential growth formula:

$$Q(a) = Q_0 + Q_1 e^{ka}, \qquad (2)$$

where $Q_0$ is the offset parameter, $Q_1$ is the initial Q-factor value, $k$ is the rate. The curve has the following parameters: $Q_0$ = -10.36, $Q_1$ = 76.38 and $k$ = 0.008.

Figure 2b shows that the Q-factor exponentially increases as the gap widens, reaching a value of $10^6$ when the gap is 1.2 μm. This behavior confirms the q-BIC nature of the resonance, as the quality factor of the q-BIC mode theoretically approaches infinity with decreasing asymmetry. The proximity of the cuboids to the 1D metasurface introduces asymmetry into the field distribution excited by the incident wave. Conversely, increasing the distance between the cuboids and the 1D metasurface restores the system's symmetry, thereby reducing radiation leakage. Such control over the Q-factor of the q-BIC resonances, realized through out-of-plane symmetry breaking, offers an additional degree of freedom for tuning their characteristics. However, Figure 1c illustrates that the transmission level decreases as the Q-factor increases, indicating greater losses at the resonance wavelength when approaching the BIC state. Furthermore, the vertical movement of the cuboids does not allow significant tuning of the central wavelength unless the separation is reduced to less than 100 nm, as shown in Figure 1c and 2a.

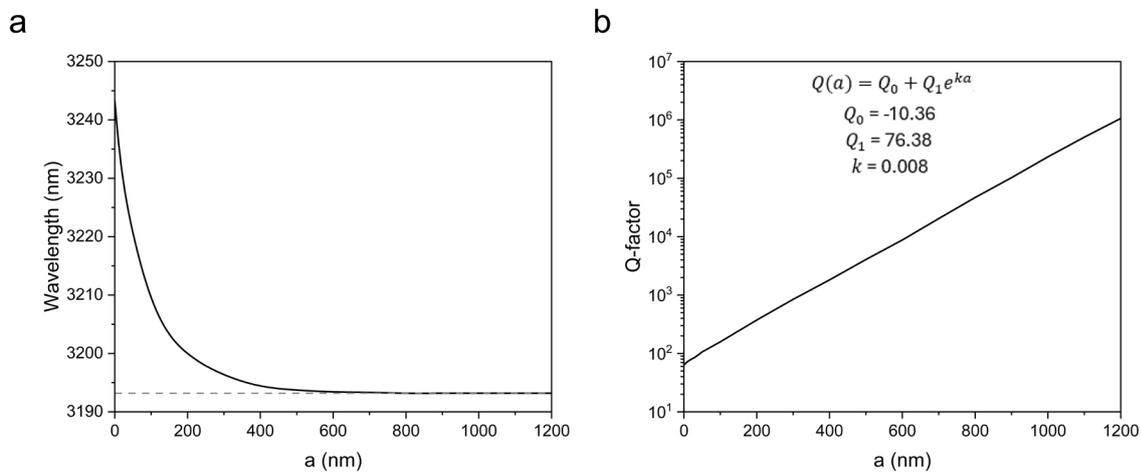

Figure 2. a, b) Dependence of the q-BIC resonance wavelength and Q-factor on the distance between the bars and cuboids.

Next, we study a more practical design containing 1D array of bars and a 2D MEMS membrane, as schematically illustrated in Figure 3a, enabling control of the same q-BIC resonance. The MEMS membrane is flat and thin, providing a significant advantage over bulk cuboids, which would need to be attached to a supporting scaffolding structure for manufacturing, complicating the process. Utilizing flat membranes instead of volumetric structures simplifies fabrication.

Figure 3a illustrates the schematic of the proposed tunable metasurface design, comprising silicon bars and a perforated MEMS membrane, along with its transmission characteristics for varying distances between the two layers (*a*). The structural parameters are as follows: $w_m$ = 0.8 µm, $h_m$ = 0.3 µm, $w_h$ = 0.7 µm, $l_h$ = 1.1 µm, $w_c$ = 0.4 µm, $w_r$ = 0.8 µm, $h_r$ = 0.25 µm, $d$ = 0.7 µm. The edge length of the square unit cell (*p*) is 1.5 µm. The polarization of the incident plane wave is shown using the triple of vectors in Figure 3a. This design enables the excitation of the q-BIC resonance, with the electric field in the bars corresponding to the loops in the bars previously depicted in Figure 1b.

Figure 3b illustrates that the vertical movement of the MEMS membrane enables control over the Q-factor of the q-BIC resonances but does not allow significant tuning of the resonance peak central wavelength. The q-BIC resonance peak wavelength shifts by less than 20 nm when tuning *a* from 0.3 to 1 µm. The height of the bars was optimized to shift the q-BIC resonance into the range of maximum reflection. This adjustment accounted for the structural differences between the cuboids and the perforated MEMS membrane, enabling out-of-band transmission of less than 0.1 within the 2600–2900 nm range when tuning *a* from 0.3 to 1 µm. The q-BIC resonance exhibits a bandwidth of approximately 1 nm when the distance between the layers (*a*) is 0.9 µm. Interestingly, the peak transmission level does not decrease with increasing the Q-factor and *a*, suggesting a more complex behavior of the q-BIC resonance compared to the previous case with cuboids and bars, where the peak transmission decreased as *a* increased. While losses remain proportional to the Q-factor, the reflection at the resonance wavelength does not increase with *a*. This observation highlights the potential for optimizing the 2D MEMS structure to maintain high transmission values at the resonance wavelength while achieving the desired ultranarrow bandwidth.

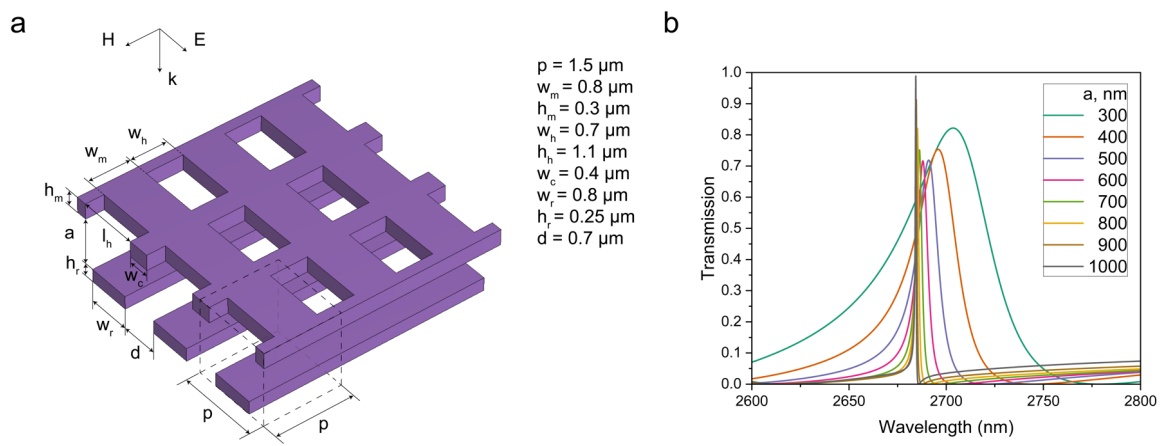

Figure 3. a) Schematic of the tunable metasurface design consisting of silicon bars and a MEMS membrane, including the unit cell indicated by the dashed lines; b) Transmission characteristics for different distances between two layers (*a*).

The bars and the overlying membrane can be attached to a supporting structure (not shown in Figure 3a). Alternatively, the bars may be placed on a quartz substrate, which can enhance the mechanical stability of the proposed metasurfaces but importantly it would increase their thickness and restrict the possibility of horizontal movement of the bars. As it was mentioned before, many tunable filter applications require maintaining a constant Q-factor while significantly shifting the resonance wavelength. To address this need, we further explore the horizontal movement of the bars to achieve additional control over the resonance peak wavelength.

Figure 4a depicts the tunable metasurface design consisting of pairs of silicon bars that can move horizontally in the direction normal to their axis, while the membrane moves vertically above the bars. The structural parameters remain consistent with those in Figure 3a, but we alter the distance between the silicon bar pairs ($d$), while keeping the period $p$ constant. The polarization of the incident plane wave is shown in Figure 4a. Two independent orthogonal tuning mechanisms enable precise control of both the central wavelength and bandwidth (or Q-factor) of the q-BIC resonance.

Figure 4b shows the transmission coefficient for various distances between the bars ($d$) when $a$ = 900 nm. The horizontal movement of the bars allows for a resonance wavelength shift of approximately 65 nm. In this case, out-of-band transmission is less than 0.1 within the 2600–3100 nm range. However, this approach alone does not maintain a constant bandwidth or Q-factor. As previously demonstrated, precise bandwidth control is achieved through the vertical movement of the membrane. By combining two independent orthogonal tuning mechanisms, complete control over the q-BIC resonance characteristics can be realized.

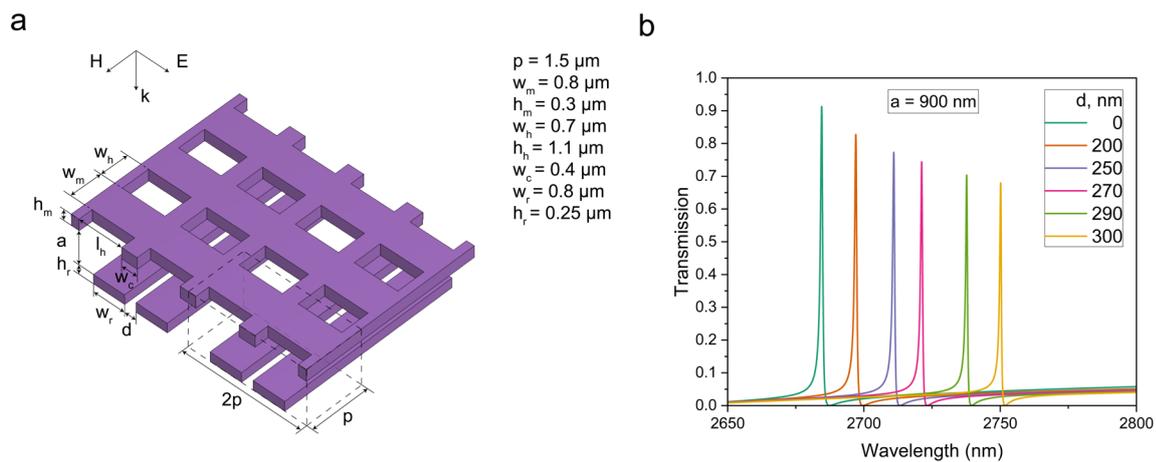

Figure 4. a) Schematic of the tunable metasurface design consisting of pairs of silicon bars and a MEMS membrane, including the unit cell indicated by the dashed lines; b) Transmission characteristics for varying distances between the bars ($d$), with the distance between the two layers ($a$) fixed at 900 nm.

Figure 5 a and b illustrates the dependence of the resonance central wavelength and Q-factor on two parameters: the distance between two layers ($a$) and the distance between the bars ($d$). By precisely adjusting both parameters, as indicated by the black lines in Figure 5b, it is possible to maintain a constant Q-factor while tuning the central wavelength by more than 60 nm.

Table 1 summarizes the characteristics of three investigated tunable metasurface designs:

(i) Metasurface concept utilizing vertically adjustable cuboids, as shown in Figure 1a;

(ii) Metasurface employing a single tuning mechanism based on a vertically adjustable perforated membrane, as illustrated in Figure 3a;
(iii) Metasurfaces incorporating two independent orthogonal tuning mechanisms, as depicted in Figure 4a.

Table 1 highlights the benefits of utilizing two independent orthogonal tuning mechanisms, which enable the preservation of a consistent Q-factor of the q-BIC resonance while substantially extending the spectral tuning range. Additionally, the design ensures flat MEMS structures and achieves out-of-band transmission of less than 0.1 across a broad range near the q-BIC resonance.

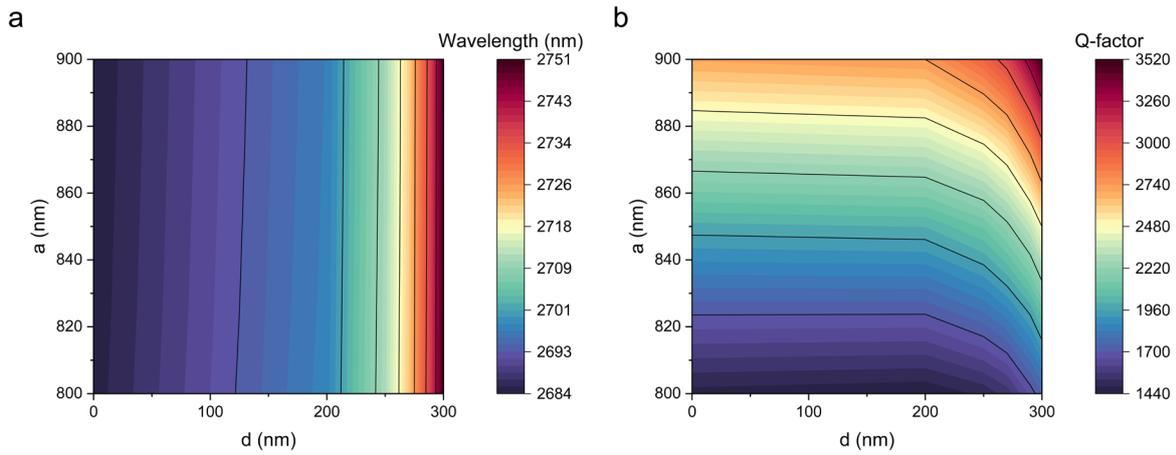

Figure 5. a, b) Dependence of the resonance central wavelength and Q-factor on two parameters: the distance between the two layers (*a*) and the distance between the bars (*d*). The black lines indicate examples of curves with constant wavelength and Q-factor, respectively.

Table 1. Comparison of characteristics of the proposed tunable metasurfaces.

| Characteristic | Metasurfaces based on vertically adjustable cuboids | Metasurfaces based on the vertically adjustable membrane | Metasurfaces based on two independent orthogonal tuning mechanisms |
|---|---|---|---|
| Spectral tuning range for *a* > 100 nm | 15 nm | 50 nm | 65 nm |
| Constant Q factor | - | - | + |
| Range in which transmission is less than 10% (for 1 nm resolution) | 2700-3500 nm | 2600-3000 nm | 2600-3100 nm |
| Flat MEMS structure | - | + | + |

**Discussion**

Our study introduces a novel degree of freedom for controlling q-BIC resonances in all-silicon metasurfaces. This is accomplished by breaking out-of-plane symmetry through the integration of a supplementary 2D structure atop a 1D metasurface where the q-BIC mode is excited. By adjusting the distance between the 2D MEMS structure and the 1D metasurface, we achieve precise control over the Q-factor of the q-BIC resonance. Furthermore, by combining two independent orthogonal tuning mechanisms within the proposed tunable metasurface designs, we demonstrated the spectral tuning range of the ultranarrow q-BIC resonance in transmission by more than 60 nm while maintaining a constant Q-factor and out-of-band transmission of less than 0.1 across a broad range.

Our theoretical work outlines pathways for the experimental realization of sub-nm resolution filters combined with a wide tuning range, enabling advancements in spectroscopy and hyperspectral imaging. The proposed tunable metasurfaces can be fabricated using techniques similar to those employed for nano-electromechanical tunable suspended gratings [30]. In this approach, a buffered silicon oxide layer beneath a 1D metasurface is partially etched to achieve suspension, while the anchors remain supported by the oxide layer, ensuring structural stability. However, we acknowledge that the experimental implementation of two independent orthogonal tuning mechanisms presents greater challenges. The bandwidth and Q-factor of the proposed metasurfaces are expected to be constrained by fabrication accuracy. Variations in the sizes of the bars, cuboids, or perforated holes in the membranes may lead to a broadening of the q-BIC resonances.

Our additional findings indicate that the horizontal movement of cuboids or membranes does not provide significant tunability of the q-BIC resonance wavelength while maintaining a 1 nm bandwidth. This suggests that achieving significant tunability of the q-BIC resonance wavelength within the studied metasurfaces requires modifications to the bars, where the maximum electric field is observed at resonance.

An alternative approach to realizing q-BIC resonances could be based on introduction of asymmetry within the meta-atom by positioning structures near one section of the meta-atom while leaving the other section unaffected by any additional layers. This method resembles in-plane symmetry breaking [35,36], where geometric asymmetry enables resonance control. However, this approach is likely to require MEMS structures to move in close proximity to 1D metasurfaces, which may pose challenges due to imperfections in the fabricated meta-structures.

Our work extends the exploration of additional degrees of freedom for manipulating the characteristics of q-BIC resonances and can be integrated with previously proposed permittivity-asymmetric methods [37] that rely on the electric control of silicon properties, as well as other approaches utilizing the thermal tuning of silicon [31]. MEMS structures that can be moved close to metasurfaces, particularly near high-symmetry points of undisturbed BIC modes, could act as perturbative elements and couplers to the radiation continuum [38]. The out-of-plane dimension can be leveraged to control optical chirality [39] and other key characteristics of metasurfaces, further enhancing the range of applications for mechanically tunable q-BIC resonances.

## Conclusion

This study presents a novel approach for controlling q-BIC resonances in all-silicon metasurfaces by utilizing mechanical movements for dynamic tunability. Through breaking out-of-plane symmetry with perforated 2D MEMS membranes integrated above 1D metasurfaces, the proposed tunable design achieves precise control over the wavelength and Q-factor of q-BIC resonances. Leveraging two independent orthogonal tuning mechanisms, the proposed method delivers a spectral tuning range exceeding 60 nm while maintaining a consistent Q-factor. This advancement holds significant promise for applications requiring tunable narrow resonances, including imaging, sensing, and spectroscopy. This dual MEMS tuning mechanism can also be applied to tune other types of resonances.

## Acknowledgements

This research was supported by the Australian Research Council Centre of Excellence for Transformative Meta-Optical Systems (Project ID CE200100010).

## References

1. C. W. Hsu, B. Zhen, A. D. Stone, J. D. Joannopoulos, and M. Soljačić, "Bound states in the continuum," Nat Rev Mater **1**, 16048 (2016).
2. K. Koshelev, A. Bogdanov, and Y. Kivshar, "Meta-optics and bound states in the continuum," Science Bulletin **64**, 836–842 (2019).
3. S. Joseph, S. Pandey, S. Sarkar, and J. Joseph, "Bound states in the continuum in resonant nanostructures: an overview of engineered materials for tailored applications," Nanophotonics **10**, 4175–4207 (2021).
4. S. I. Azzam and A. V. Kildishev, "Photonic Bound States in the Continuum: From Basics to Applications," Adv. Optical Mater. **9**, 2001469 (2021).
5. Y. Zeng, X. Zhang, X. Ouyang, Y. Li, C. Qiu, Q. Song, and S. Xiao, "Manipulating Light with Bound States in the Continuum: from Passive to Active Systems," Advanced Optical Materials **12**, 2400296 (2024).
6. K. Fan, I. V. Shadrivov, and W. J. Padilla, "Dynamic bound states in the continuum," Optica **6**, 169 (2019).
7. T. Gu, H. J. Kim, C. Rivero-Baleine, and J. Hu, "Reconfigurable metasurfaces towards commercial success," Nat. Photon. **17**, 48–58 (2023).
8. T. Hu, Z. Qin, H. Chen, Z. Chen, F. Xu, and Z. Wang, "High-Q filtering and dynamic modulation in all-dielectric metasurfaces induced by quasi-BIC," Opt. Express **30**, 18264 (2022).
9. F. V. Kovalev, A. E. Miroshnichenko, A. A. Basharin, H. Toepfer, and I. V. Shadrivov, "Active Control of Bound States in the Continuum in Toroidal Metasurfaces," Advanced Photonics Research 2400070 (2024).
10. X. Chen, W. Fan, and H. Yan, "Toroidal dipole bound states in the continuum metasurfaces for terahertz nanofilm sensing," Opt. Express **28**, 17102 (2020).

11. W. Cen, T. Lang, Z. Hong, J. Liu, M. Xiao, J. Zhang, and Z. Yu, "Ultrasensitive Flexible Terahertz Plasmonic Metasurface Sensor Based on Bound States in the Continuum," IEEE Sensors J. **22**, 12838–12845 (2022).
12. Z. Jing, W. Jiaxian, G. Lizhen, and Q. Weibin, "High-Sensitivity Sensing in All-Dielectric Metasurface Driven by Quasi-Bound States in the Continuum," Nanomaterials **13**, 505 (2023).
13. Y. Xu, L. Zhang, B. Du, H. Chen, Y. Hou, T. Li, J. Mao, and Y. Zhang, "Quasi-BIC Based Low-Voltage Phase Modulation on Lithium Niobite Metasurface," IEEE Photon. Technol. Lett. **34**, 1077–1080 (2022).
14. Z. Yang, M. Liu, D. Smirnova, A. Komar, M. Shcherbakov, T. Pertsch, and D. Neshev, "Ultrafast Q-boosting in semiconductor metasurfaces," Nanophotonics **13**, 2173–2182 (2024).
15. H. Zheng, Y. Zheng, M. Ouyang, H. Fan, Q. Dai, H. Liu, and L. Wu, "Electromagnetically induced transparency enabled by quasi-bound states in the continuum modulated by epsilon-near-zero materials," Opt. Express **32**, 7318 (2024).
16. G. Xu, H. Xing, Z. Xue, D. Lu, J. Fan, J. Fan, P. P. Shum, and L. Cong, "Recent Advances and Perspective of Photonic Bound States in the Continuum," Ultrafast Sci **3**, 0033 (2023).
17. H. B. Khaniki, M. H. Ghayesh, and M. Amabili, "A review on the statics and dynamics of electrically actuated nano and micro structures," International Journal of Non-Linear Mechanics **129**, 103658 (2021).
18. S. Nazir and O. S. Kwon, "Micro-Electromechanical Systems-based Sensors and Their Applications," Appl. Sci. Converg. Technol. **31**, 40–45 (2022).
19. M. Martyniuk, K. K. M. B. D. Silva, G. Putrino, H. Kala, D. K. Tripathi, G. Singh Gill, and L. Faraone, "Optical Microelectromechanical Systems Technologies for Spectrally Adaptive Sensing and Imaging," Adv Funct Materials **32**, 2103153 (2022).
20. Y. Ma, W. Liu, X. Liu, N. Wang, and H. Zhang, "Review of sensing and actuation technologies – from optical MEMS and nanophotonics to photonic nanosystems," International Journal of Optomechatronics **18**, 2342279 (2024).
21. A. L. Holsteen, A. F. Cihan, and M. L. Brongersma, "Temporal color mixing and dynamic beam shaping with silicon metasurfaces," Science **365**, 257–260 (2019).
22. Z. Ren, Y. Chang, Y. Ma, K. Shih, B. Dong, and C. Lee, "Leveraging of MEMS Technologies for Optical Metamaterials Applications," Adv. Optical Mater. **8**, 1900653 (2020).
23. Y. Chang, J. Wei, and C. Lee, "Metamaterials – from fundamentals and MEMS tuning mechanisms to applications," Nanophotonics **9**, 3049–3070 (2020).
24. N. I. Zheludev and E. Plum, "Reconfigurable nanomechanical photonic metamaterials," Nature Nanotech **11**, 16–22 (2016).
25. Y. Zhao, Z. Liu, C. Li, W. Jiao, S. Jiang, X. Li, J. Duan, and J. Li, "Mechanically reconfigurable metasurfaces: fabrications and applications," npj Nanophoton. **1**, 16 (2024).
26. R. Magnusson and Y. Ding, "MEMS tunable resonant leaky mode filters," IEEE Photon. Technol. Lett. **18**, 1479–1481 (2006).
27. T. Sang, T. Cai, S. Cai, and Z. Wang, "Tunable transmission filters based on double-subwavelength periodic membrane structures with an air gap," J. Opt. **13**, 125706 (2011).
28. G. Quaranta, G. Basset, O. J. F. Martin, and B. Gallinet, "Recent Advances in Resonant Waveguide Gratings," Laser & Photonics Reviews **12**, 1800017 (2018).
29. Y. H. Ko, K. J. Lee, S. Das, N. Gupta, and R. Magnusson, "Micro-electromechanical-system-tuned resonant filters spanning the 8–12 μm band," Opt. Lett. **46**, 1329 (2021).
30. H. Kwon, T. Zheng, and A. Faraon, "Nano-electromechanical Tuning of Dual-Mode Resonant Dielectric Metasurfaces for Dynamic Amplitude and Phase Modulation," Nano Lett. **21**, 2817–2823 (2021).

31. X. Zhou, X. Zhang, Y. Zuo, Z. Chen, Z. Qian, Z. Zhang, C. Peng, and H. Li, "Ultra-compact on-chip spectrometer based on thermally tuned topological miniaturized bound states in the continuum cavity," Opt. Lett. **48**, 4993 (2023).
32. F. Tang, J. Wu, T. Albrow-Owen, H. Cui, F. Chen, Y. Shi, L. Zou, J. Chen, X. Guo, Y. Sun, J. Luo, B. Ju, J. Huang, S. Liu, B. Li, L. Yang, E. A. Munro, W. Zheng, H. J. Joyce, H. Chen, L. Che, S. Dong, Z. Sun, T. Hasan, X. Ye, Y. Yang, and Z. Yang, "Metasurface spectrometers beyond resolution-sensitivity constraints," Sci. Adv. **10**, eadr7155 (2024).
33. M. Galli, S. L. Portalupi, M. Belotti, L. C. Andreani, L. O'Faolain, and T. F. Krauss, "Light scattering and Fano resonances in high-Q photonic crystal nanocavities," Applied Physics Letters **94**, 071101 (2009).
34. A. E. Miroshnichenko, S. Flach, and Y. S. Kivshar, "Fano resonances in nanoscale structures," Rev. Mod. Phys. **82**, 2257–2298 (2010).
35. Z. Zheng, A. Komar, K. Zangeneh Kamali, J. Noble, L. Whichello, A. E. Miroshnichenko, M. Rahmani, D. N. Neshev, and L. Xu, "Planar narrow bandpass filter based on Si resonant metasurface," Journal of Applied Physics **130**, 053105 (2021).
36. C. Zhou, L. Huang, R. Jin, L. Xu, G. Li, M. Rahmani, X. Chen, W. Lu, and A. E. Miroshnichenko, "Bound States in the Continuum in Asymmetric Dielectric Metasurfaces," Laser & Photonics Reviews **17**, 2200564 (2023).
37. R. Berté, T. Weber, L. De Souza Menezes, L. Kühner, A. Aigner, M. Barkey, F. J. Wendisch, Y. Kivshar, A. Tittl, and S. A. Maier, "Permittivity-Asymmetric *Quasi* -Bound States in the Continuum," Nano Lett. **23**, 2651–2658 (2023).
38. G. Q. Moretti, A. Tittl, E. Cortés, S. A. Maier, A. V. Bragas, and G. Grinblat, "Introducing a Symmetry-Breaking Coupler into a Dielectric Metasurface Enables Robust High-Q Quasi-BICs," Advanced Photonics Research **3**, 2200111 (2022).
39. L. Kühner, F. J. Wendisch, A. A. Antonov, J. Bürger, L. Hüttenhofer, L. De S. Menezes, S. A. Maier, M. V. Gorkunov, Y. Kivshar, and A. Tittl, "Unlocking the out-of-plane dimension for photonic bound states in the continuum to achieve maximum optical chirality," Light Sci Appl **12**, 250 (2023).